\documentclass{sf2a-conf2024}
\usepackage{graphicx}
\usepackage{hyperref}
\usepackage[]{natbib}  
\usepackage{epstopdf}

\def\BibTeX{{\rm B\kern-.05em{\sc i\kern-.025em b}\kern-.08em
    T\kern-.1667em\lower.7ex\hbox{E}\kern-.125emX}}
\bibpunct{(}{)}{;}{a}{}{,}  %%%%%%%%%%%%%  A&A bibliography style
\begin{document}

\TitreGlobal{SF2A 2024}

%%-----------------------------------------------------------------
%%      the top matter
%%

\title{On the links between astronomy, \\astronomers and Science Fiction}

\runningtitle{Astronomy and Science Fiction}

\author{S. Boissier}\address{Aix Marseille Univ, CNRS, CNES, LAM, Marseille, France}

%% Keep this line, even if the page will be settled afterwards.
\setcounter{page}{237}

%%-----------------------------------------------------------------

\maketitle

%%-----------------------------------------------------------------
%%        The abstract
%% 
%%  Warning!  within the abstract:
%%  - do not use macros. 
%%  - do not use commands like: \cite, \citet, \citep ... etc.

\begin{abstract}
Science Fiction is using astronomy to offer to the public blockbusters at the movies (e.g. Interstellar), series or movies in streaming media (Don’t Look Up, The Expanse), many books from classic authors (I. Asimov, A.C. Clarke) or more moderns ones (K.S. Robinson), comics (the adventures of Valerian and Laureline), or video games (Mass Effect, No Man's sky) that have a very large cumulated audience.

Astronomers can use Science Fiction to illustrate physics or astronomical facts. It might be a good way to talk about our work, our methods, by comparing them to examples with which a large audience is familiar. A few examples are provided in this contribution.

In a recent study (Stanwey, 2022), it was shown that 93 percent of British professional astronomers have an interest for Science Fiction, and 69 percent consider that Science Fiction influenced their career or life choice. I am presenting a similar study made for French astronomers, performed during and just after the 2024  French astronomer meeting (Journées de la SF2A).
\end{abstract}
%
%
%% Insert the keywords (to appear in the ADS indexing)
%% Keywords must be separated by a comma

%%-----------------------------------------------------------------

\section{Introduction}
%%---------------------

In a time when fake news and obscurantism knock at our doors, and science is sometimes considered as
an opinion without any additional value than any other one, it is important to bring outreach not only to astronomy 
aficionados but also to those who may believe they do not have the baggage to 
follow an astronomy conference, those who may think the will feel lost in technological lingo, or 
even those who have hostile opinions about scientists. 
%To touch this public, we need other angles of approach. 
We need to find new ways to engage with this public.
Fortunately, it is possible to talk about science, astronomy in particular, while discussing 
about art \citep[e.g.][]{olson2014,luminet2023} or Science Fiction
\citep[e.g.][and references therein]{ivanov2023,orthia2019}.

 \section{Astronomy as a source of inspiration for science-fiction}
 
Science Fiction (in a very broad sense) tells fantastic and marvellous stories, usually based on the possible development of science or mankind, and often linked to space. The Science Fiction stories related to space are numerous, and share the sky with astronomers \citep[e.g.][]{peron2015}. The stars and planets have forever inspired writers, back e.g. to Lucien de Samosate in the second century, and later Cyrano de Bergerac (17th century) with their travels to the Sun and the Moon. 
In 1752, Voltaire wrote ``Micromégas'', a text that clearly takes into account the astronomical knowledge of the time, while Maupassant wrote ``L'Homme de Mars'' (the Man from Mars) in 1887, inspired by recent observations of the red planet.

Nowadays, several Science Fiction writers clearly show a very good astronomical knowledge, and impregnate their stories with it, what is especially true in what is sometimes called ``hard Science Fiction''. Among them, Kim Stanley Robinson describes Mars almost with an atlas precision. Alastair Reynolds put his characters around real stars near Earth and takes into account the limits of space travel \citep[I must admit I read his Revelation space series, but not his PhD, ][]{reynolds1992}.
Like him, many Science Fiction writers have had a very close connection with astronomy, some are actual astronomers (for instance Fred Hoyle -who coined the 
``big bang'' term, or Sarah Anderson -member of the LOC of this meeting), and others are acclaimed for their knowledge of astronomy/space science/physics (e.g. Arthur C. Clarke, Greg Egan). 
Often, also, astronomers are not the creators of a piece, but are advisors.
%
%Sometimes, creative peoples are directly advised by scientists. 
In the Futurama TV show, some astronomical Easter eggs come from David Schiminovich from Columbia university (as we can learn by listening to the comments track on one of the DVDs).

Clearly our field of science inspires writers and artists, some of them making adequate depictions of the sky, planets, stars and galaxies, while other take some liberties with reality, for the sake of storytelling often, by ignorance sometimes.

\section{Using Science Fiction as a pedagogical resource }
%%-------------------------

It is largely recognized that Science Fiction has been used in research in a large range of disciplines, and especially with the goal of teaching, outreach, and favouring scientific careers \citep[e.g.][]{menadue2017}.

In France, Roland Lehoucq made a speciality of speaking of the law of physics based on science-fiction. He is the author of many public conferences and books based on this principle \citep[e.g.][]{lehoucq11}. In ``Voyages dans le futur'', \citet{prantzos1998} explores the future of space exploration, and the long-term destiny of the Universe, with an extremely rich list of references including technical reports, prospective works, and science-fiction novels.

I am embracing this idea that we can talk about science on a fictional basis, to reach new audiences, and I will present below a few personal examples.

I had recently the opportunity to participate to the colloquium ``Quand la Science Fiction change le monde''\footnote{\url{https://espritfutur.fr/quand-la-science-fiction-change-le-monde/}} in Aix En Provence in April 2024, with authors, editors, and academics from the humanities. With Franck Selsis, I was representing astronomy at this event, meeting with persons that feel disconnected from ``hard'' science. This is an example of interdisciplinary event that allows us to enlarge our audience.

In \citet{boissierperon}, we shared with the French audience the fact that Tolkien was fond of astronomy, what can be observed in ``Lord of the Ring'' or ``The hobbit'', as clearly established by Kristine Karsen\footnote{Her website includes an impressive collection of links of studies and documents:  \url{https://www.physics.ccsu.edu/larsen/tolkien.html}} in her numerous publications and works on the subject \citep[e.g.][]{larsen2004}. Although Tolkien is not mainstream science-fiction, we can fairly say it touches a broad audience on the basis of the marvellous. 
A similar work could be done with Lovecraft, the master of horror stories. The French review ``Ciel \& Espace'' dedicated an article to his very strong interest for astronomy, his observations at Ladd observatory (it is actually possible to browse through his  observations notebook\footnote{\url{https://digital.library.villanova.edu/Item/vudl:591771}}), his writings about astronomy, and the influence of our field into his stories  (Ciel \& Espace 561, 74).
In Lovecraft, the source of the greatest horror has literally fallen from the sky! The direct link between his texts and astronomy would deserve some more words to discuss if the gibbous moon was his favourite source of light, as advanced by \citet{houellebecq1991}, which stars are found in his stories (e.g. Algol, Polaris), and how his depiction of astronomy is related to his own observations, and to the astronomical state of the art of the time.

With the large audience of public conferences, I found it  pleasant to discuss about Science Fiction and astronomy. It is easy to find examples allowing to describe the work of a scientist, the nature of science, the facts of astronomy, in a very didactic and fun way, by contrasting what is shown in blockbusters or in books to reality.
One very interesting example is the depiction of astronomers. For instance, in the movie ``Don't Look Up'', despite some correct incarnations and adequate presentations of some aspects, the lights stay on in the dome of the telescope during observations, and Leonardo di Caprio needed a hand double to write some basic equations! Other descriptions of astronomers in fiction can be found in \citet{crovisier2011} for Jules Verne novels. Despite some caricatures \citep{west2011}, a lot can be said about astronomers and astronomy based on fictional astronomers (I have a personal knowledge of about 40 fictional astronomers, so there is the material to speak about a lot of things!).
One of the aspect of this type of use of Science Fiction is the promotion of science and astronomy career. Even if characters are often caricatured, they inspire young students, and especially women. Samantha Carter in the TV series Stargate is portrayed as a bright astrophysicist (in addition to a military explorer). She is quite an unrealistic scientist. Nevertheless  a young astronomer told me she got inspired by this strong character living adventure at the surface of what should be called exo-planets !
More widely discussed is the influence of the character Dana Scully (X-files) on the number of girls embracing studies in science, medicine, engineering (the so called ``Scully 
Effect''\footnote{\url{https://en.wikipedia.org/wiki/Dana_Scully}}).
Similarly, a ``Uhura effect'' has been proposed based on the character from Star Trek, and her popularity has 
been used for various causes, including by NASA to encourage minorities to join the organisation. While these effects are popular on anecdotal basis, they are still under studies (see \url{https://sites.google.com/neiu.edu/theuhuraeffect-dwatson/home}). More generally, the identification of young persons with scientists in fiction is the subject of many studies \citep[see][ans references therein]{steinke12}.

\begin{figure}[ht!]
 \centering
 \includegraphics[width=0.48\textwidth,clip]{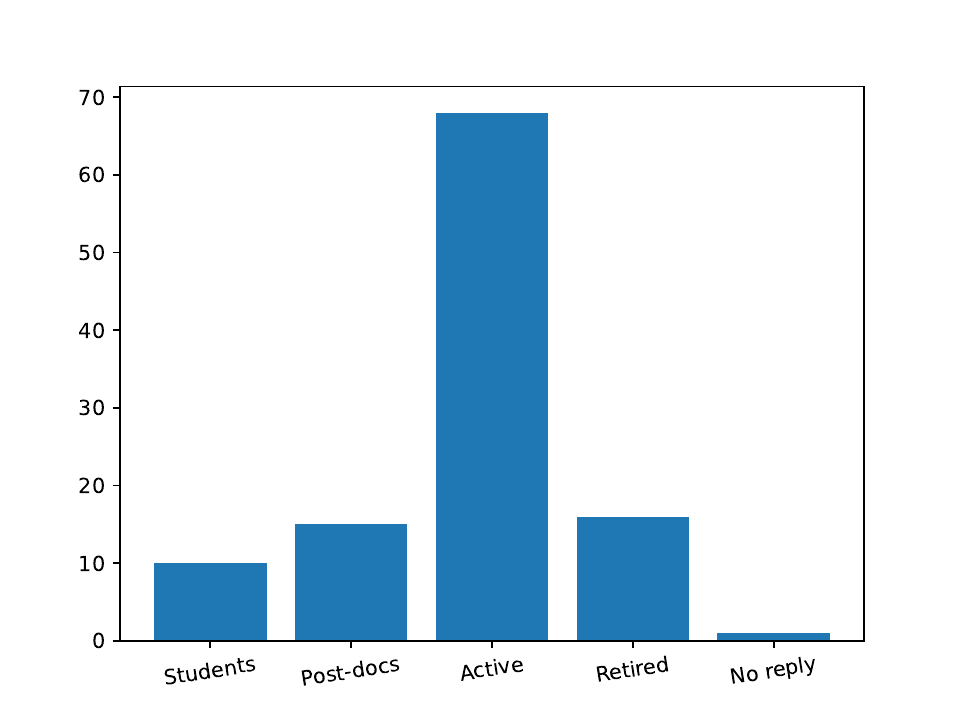}%      
 \includegraphics[width=0.48\textwidth,clip]{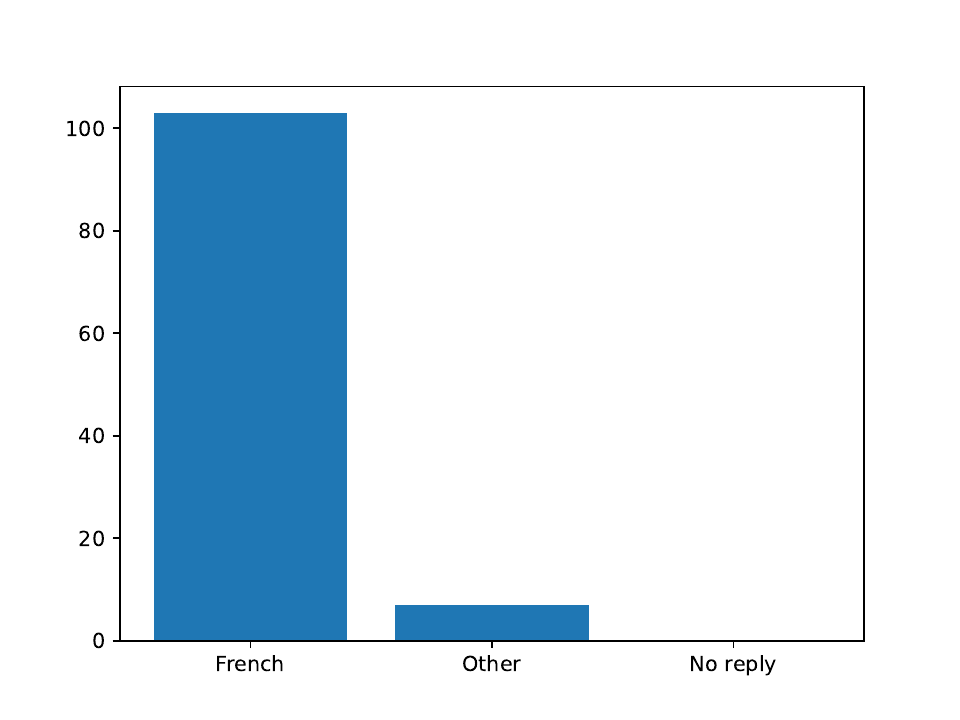}      
%% Note the ABSENCE of the extension .pdf  !
  \caption{Categories (left) and Nationalities  (right) of the persons who replied the poll.}
  \label{BoissierS09:fig1}
\end{figure}

\section{A poll proposed to the SF2A 2024 meeting participants}

\subsection{Rationale and methodology}

In this last section, I will present the results of a poll that was proposed at the SF2A meeting, inspired by the work of \citet{stanway22}. She made an analysis of the British astronomers and their link to Science Fiction. 
The most striking of her results was the statement that 93 percent of British astronomers have an interest in Science Fiction, and that 69 percent consider that this taste influenced their career choice or life. Other studies concerning the possible influence of career science can be found \citep{orthia2019} but are not limited to astronomy or astronomers.

My gut feeling was that my French colleagues would not be that much interested, and that few of them  would admit to have been influenced by Science Fiction. But science teach us not to listen to our gut feelings, thus I decided to do a similar analysis as 
\citet{stanway22} but directed to the French astronomy community. By design, I thus decided to ask very similar questions, although the methodology is slightly different.
\citet{stanway22} polled the participants to a National Astronomy Meeting (NAM) of the British community. She presented an interactive poster in which participants could indicate their love/loath of Science Fiction and its possible influence by putting a sticker on a 2D plot similar to Fig 3. For quantitative assessment, her axis were digitized between -1 and 1, but some participants put stickers beyond the limits of the plot.
I wanted to poll the participants of the French Astronomical meeting (Journées de la SF2A 2024), but I could not follow the same procedure because there was no physical poster! Thus, I created an online form, asking similar questions (interest in Science Fiction from 1 to 10, influence on their life or career from 1 to 10). I included a few more optional questions on the participants (status, nationality) to eventually distinguish various populations and be able to have a ``French astronomer'' sample. To put our results on the same axis, I converted her -1 to 1 scale to my 1 to 10 scale with a linear transformation (neglecting the fact that some answers in her survey were beyond the limits, but it is hard to find another meaningful transformation). During the meeting, this form was advertised during the session on pedagogy. Finally, the form link was sent in the subsequent newsletter of the SF2A (the astronomy French society).

\subsection{The participants}

With about 500 participants to the meeting, and about 2000 persons who are registered to the mailing list, the number of answers to the poll remains quite small: only 110 answers. This is to compare to the \citet{stanway22} 239 answers for 643 attendees to the meeting. My results are thus very qualitative and incomplete.
Fig. \ref{BoissierS09:fig1} shows the demography of the persons who replied : mostly French, professional and active astronomers. In the following, I tried to split my results according to the different status, but due to small numbers, they are indistinguishable.

\subsection{Results}

\begin{figure}[ht!]
 \centering
 \includegraphics[width=0.48\textwidth,clip]{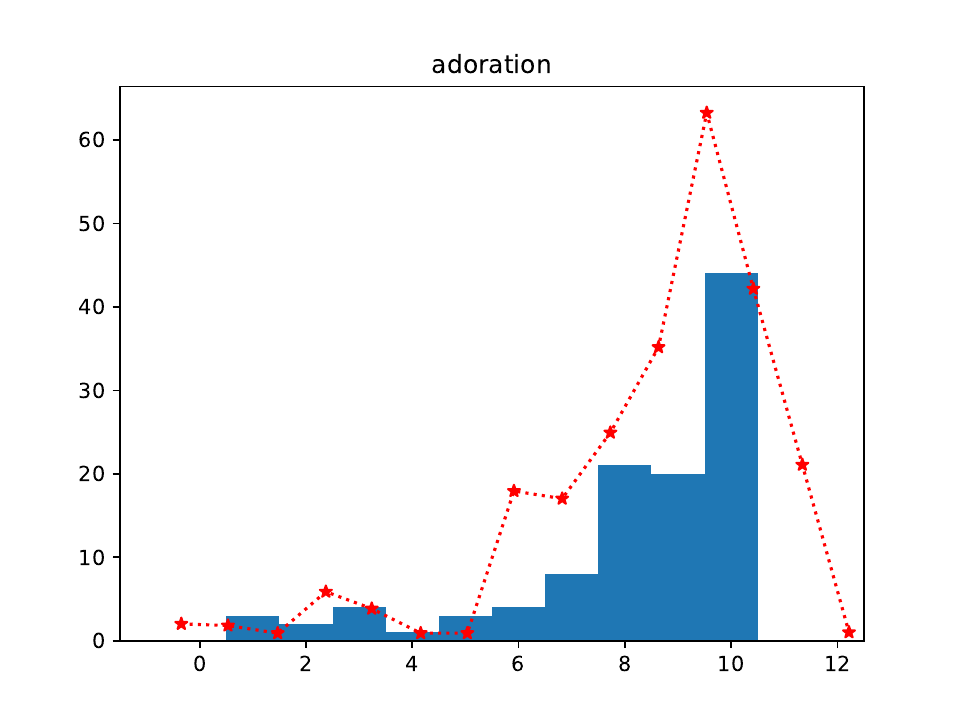}%      
 \includegraphics[width=0.48\textwidth,clip]{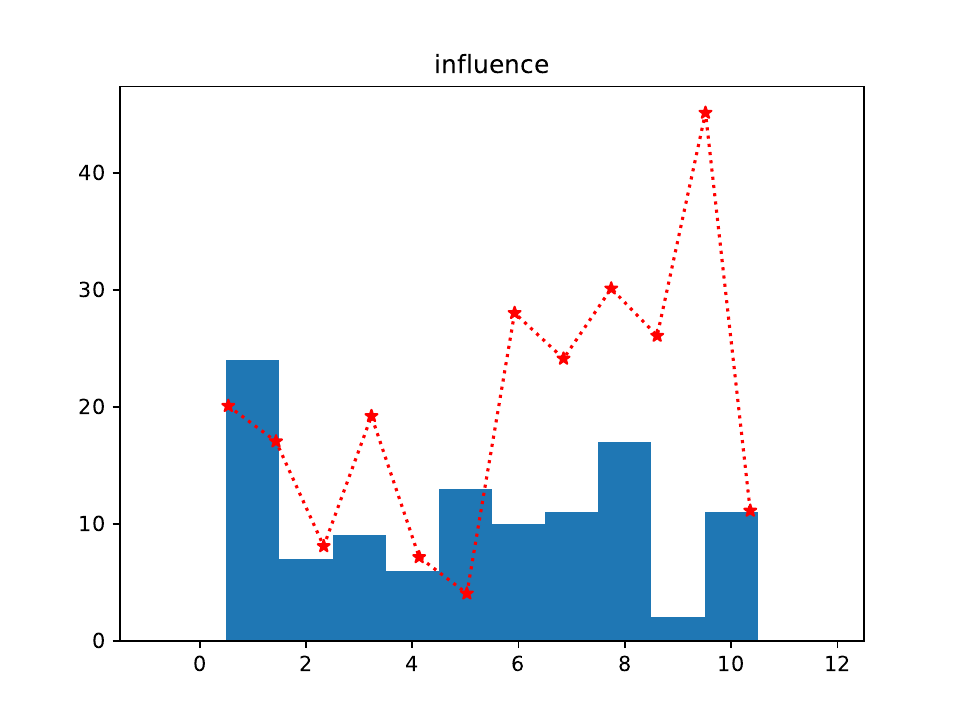}      
%% Note the ABSENCE of the extension .pdf  !
  \caption{Adoration (left) and Influence (right). The stars show the histograms 
  from \citet{stanway22}, with the x-axis converted from her -1 to 1 range to my 1 to 10 range.}
  \label{BoissierS09:fig2}
\end{figure}

Fig. \ref{BoissierS09:fig2} shows the histograms of ``adoration'', and ``influence'' of science-fiction on my 1 to 10 scale. This is to be compared to the loathe/love axis and influenced/not influenced axis from \citet{stanway22} on her -1 to 1 scale.
Despite the small sample, it is visible that the ``adoration'' of French astronomers (at least the ones who replied) is similar in shape to the adoration of the British one, with a clear peak close to the maximal value of the poll. 
On the contrary, the ``influence'' histogram presents a clear difference. \citet{stanway22} found a distribution favouring the large influence range with respect to my study for which it is rather flat. 

\begin{figure}[ht!]
 \centering
 \includegraphics[width=0.8\textwidth,clip]{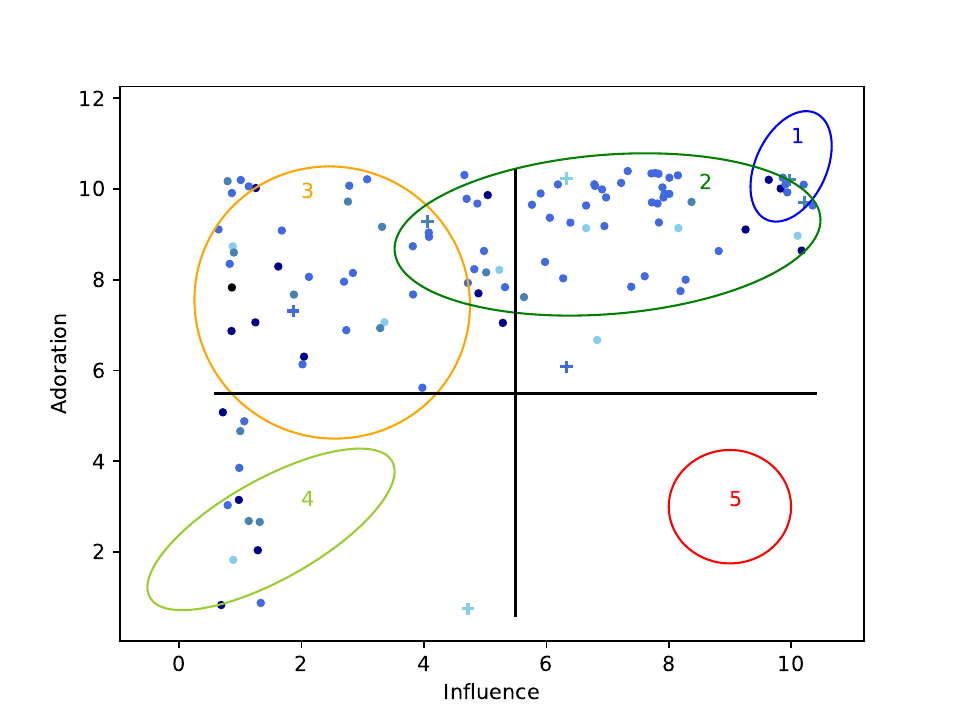}    
%% Note the ABSENCE of the extension .pdf  !
  \caption{2D distribution of the answers, adding a random value between -0.4 and 0.4 to each point on each axis to distinguish them. Darker colours indicate more advanced steps of the career (students, post-docs, active, retired), and ``+'' symbols indicate foreigners. The ellipses indicate the regions defined in \citet{stanway22}, see text of section \ref{Boissier:sec2D}.}
  \label{BoissierS09:fig3}
\end{figure}

To further compare this study and \citet{stanway22}, Fig. \ref{BoissierS09:fig3} shows the 2D distributions of the results. The number of points in each quadrant is given in table \ref{BoissierS09:tab1} for both studies.
Like their British colleagues, The French astronomers (who replied) like science-fiction very much. The fraction that dislikes science-fiction is however slightly higher in France. Taking into account the fact that the answers concerns only 20\%  of the participants at the meeting, the 88\% of French astronomer loving Science Fiction might be a large overestimation with respect to the full population. In the British case, \citet{stanway22} has a larger fraction of SF lovers, and a larger participation (one third of the meeting participants) and argues against the biases (not everyone at the meeting could see the poster, independently of their like/dislike of Science Fiction). Thus, I cannot exclude that the fraction of SF lover in the UK is larger than in France.

Concerning the influence of Science Fiction on career and life choices, the difference is striking. British astronomers loving Science Fiction are much more inclined to recognize an influence.
I do not have a solid interpretation for this fact, but I am advancing the hypothesis that Science Fiction (and all that is related to the marvellous) is just not considered as a very serious field of literature (or art) in France. It is often relegated to a sub-genre interesting teenagers and geeks. Despite the strong historical influence of Jules Verne, no authors in France are as much celebrated and recognized as their anglo-saxon counterparts (e.g. Tolkien, Douglas Adams, or Arthur C. Clarke), and Science Fiction is rarely considered seriously in the academic world \citep[see however][]{brean2012}.

\begin{table}
\begin{tabular}{| l |l | l | }
\hline
                               & French (this study)   & British \citep{stanway22} \\
\hline
Love SF and influenced         &  46\% (51/110)    & 66\% (157/239) \\
Love SF but not influenced     &  42\% (46/110)    & 27	\% (66.239) \\			
Disklike SF and not influenced &  12\% (13/110)      &  4  \% (9/239) \\	
Dislike SF but influenced      &  0\% (0/110)      &  3  \%  (7/239) \\ 
\hline
\end{tabular}
\caption{\label{BoissierS09:tab1}Statistics in the quadrant of the adoration/influence distribution of Fig. 3 for this study and \citet{stanway22}.}
\end{table}

In Fig. \ref{BoissierS09:fig3} we also reproduce the 5 regions that \citet{stanway22} humorously defined:

\label{Boissier:sec2D}
1) The assymptotic SF-lovers branch

Those are the astronomers that are at the extrema of SF adoration and influence. In the case of \citet{stanway22}, participants could go beyond the axis, and it was clearly identified. In the French astronomer case, it is less prominent by construction, but we do find this type of answers close to both extrema.

2) The main sequence

Astronomers that love Science Fiction, and are increasingly influenced by it define a sequence of 
increasing adoration with influence, that is also visible in this study.

3)The weakly interacting cloud

Astronomers who love Science Fiction with various range, and are not much influenced. I find this structure in this survey too.

4) The Science Fiction Haters cooling track.

The trend found by \citet{stanway22} of smaller influence with increasing hate (low values of adoration) is not found here. Instead French haters all indicated all a very small level of influence.

5) The D clump

Astronomers that expressed a strong influence of Science Fiction on their choice / careers despite not loving Science Fiction. This clump is just inexistent in our data. This clump was actually surprising to find and hard to explain. Maybe French astronomers are more rational than British ones. As for the other small differences above, maybe it should be related to the different methodology applied in both studies.

%\subsection{Hyperlinks}
%Hyperlinks can be introduced as follows: \url{http://www.sf2a.eu/}.

%%
%% Example of single figure
%%

\section{Conclusions}
%%--------------------

Astronomy and Science Fiction are clearly related. Both astronomers, and Science Fiction lovers could sing along the lyrics of the Ballad of Serenity: ``You can't take the sky from me''. It is possible to use Science Fiction to talk about astronomy and science to a public that may not be met with other forms of outreach.

On a sociological note, by performing a survey of French astronomers, this contribution shows that an important fraction of French astronomers love Science Fiction, but this fraction is probably smaller than in the British community. The difference is even stronger for the influence of Science Fiction on career and life choices of astronomers. French astronomers are much less inclined to recognize it. I suggest this is directly related to the fact that French people do not take Science Fiction seriously enough (but I might be biased).

% Optional acknowledgements
% -------------------------
\begin{acknowledgements}
I would like to thank the organizers for giving me the opportunity to talk at the S09 session of the SF2A 2024 meeting, the SOC of the meeting. Thank you to my colleagues and friends with whom I often talked about original outreach approaches (S. Vives in particular), and science-fiction (F. Madec in particular). And another thank you to Caroline Péron with whom I share this interest for Science Fiction (among many things), and who proofread this document.
\end{acknowledgements}

\bibliographystyle{aa}  % A&A bibliography style file (aa.bst)
\bibliography{Boissier_S09} % your references in file: Yourfile.bib

\begin{thebibliography}{17}
\expandafter\ifx\csname natexlab\endcsname\relax\def\natexlab#1{#1}\fi

\bibitem[{{Boissier} \& {Péron}(2020)}]{boissierperon}
{Boissier}, S. \& {Péron}, C. 2020, Cahier Clairaut, 169, 123

\bibitem[{Bréan(2012)}]{brean2012}
Bréan, S. 2012, Science fiction en France. Théorie et histoire d'une
  littérature (SUP)

\bibitem[{{Crovisier}(2011)}]{crovisier2011}
{Crovisier}, J. 2011, in The Role of Astronomy in Society and Culture, ed.
  D.~{Valls-Gabaud} \& A.~{Boksenberg}, Vol. 260, 321--326

\bibitem[{Houellebecq(1991)}]{houellebecq1991}
Houellebecq, M. 1991, H. P. Lovecraft. Contre le monde, contre la vie (Rocher
  Eds Du)

\bibitem[{Ivanov(2023)}]{ivanov2023}
Ivanov, V.~D. 2023, Popular astronomy and other science articles in glossy
  magazines -- outreaching to those who do not care to be reached

\bibitem[{{Larsen}(2004)}]{larsen2004}
{Larsen}, K. 2004, in APS Meeting Abstracts, Vol. 2004, APS March Meeting
  Abstracts, L27.006

\bibitem[{Lehoucq(2011)}]{lehoucq11}
Lehoucq, R. 2011, SF: la science mène l'enquête (Le Pommier)

\bibitem[{Luminet(2023)}]{luminet2023}
Luminet, J.-P. 2023, Les nuits étoilées de Vincent Van Gogh (Seghers)

\bibitem[{Menadue \& Cheer(2017)}]{menadue2017}
Menadue, C.~B. \& Cheer, K.~D. 2017, Sage Open, 7, 2158244017723690, publisher:
  SAGE Publications

\bibitem[{Olson(2014)}]{olson2014}
Olson, D. 2014, Celestial Sleuth. Using Astronomy to Solve Mysteries in Art,
  History and Literature (Springer)

\bibitem[{Orthia(2019)}]{orthia2019}
Orthia, L.~A. 2019, Journal of Science Communication, 18, A08, publisher: SISSA
  Medialab srl

\bibitem[{Prantzos(1998)}]{prantzos1998}
Prantzos, N. 1998, Voyages dans le futur (Seuil)

\bibitem[{Péron(2015)}]{peron2015}
Péron, C. 2015, Criée aux livres: astronomie et création,
  \url{https://www.calameo.com/books/0039589247e5f33d0f117}

\bibitem[{{Reynolds}(1992)}]{reynolds1992}
{Reynolds}, A.~P. 1992, PhD thesis, Saint Andrews University, UK

\bibitem[{{Stanway}(2022)}]{stanway22}
{Stanway}, E.~R. 2022, arXiv e-prints, arXiv:2208.05825

\bibitem[{Steinke {et~al.}(2012)Steinke, Applegate, Lapinski, Ryan, \&
  Long}]{steinke12}
Steinke, J., Applegate, B., Lapinski, M., Ryan, L., \& Long, M. 2012, Science
  Communication, 34, 163, publisher: SAGE Publications Inc

\bibitem[{{West}(2011)}]{west2011}
{West}, M.~J. 2011, in The Role of Astronomy in Society and Culture, ed.
  D.~{Valls-Gabaud} \& A.~{Boksenberg}, Vol. 260, 411--419

\end{thebibliography}

\end{document}